# Cryogenic packaging of nanophotonic devices with a low coupling loss < 1 dB

Beibei Zeng,[1,a)] Chawina De-Eknamkul,[1,a)] Daniel Assumpcao,[2] Dylan Renaud,[2] Zhuoxian Wang,[1] Daniel Riedel,[1] Jeonghoon Ha,[1] Carsten Robens,[1] David Levonian,[1] Mikhail Lukin,[3] Ralf Riedinger,[4] Mihir Bhaskar,[1] Denis Sukachev,[1] Marko Loncar,[2] Bart Machielse[1,b)]

[1]*AWS Center for Quantum Networking, Boston, MA 02135, USA*

[2]*John A. Paulson School of Engineering and Applied Sciences, Harvard University, Cambridge, MA 02138, USA*

[3]*Department of Physics, Harvard University, Cambridge, MA 02138, USA*

[4]*Institut für Laserphysik und Zentrum für Optische Quantentechnologien, Universität Hamburg, 22761 Hamburg, Germany*

Robust, low-loss photonic packaging of on-chip nanophotonic circuits is a key enabling technology for the deployment of integrated photonics in a variety of classical and quantum technologies including optical communications and quantum communications, sensing, and transduction. To date, no process has been established that enables permanent, broadband, and cryogenically-compatible coupling with sub-dB losses from optical fibers to nanophotonic circuits. Here we report a technique for reproducibly generating a permanently packaged interface between a tapered optical fiber and nanophotonic devices with a record-low coupling loss < 1 dB per facet at near-infrared wavelengths (~730 nm) that remains stable from 300 K to 30 mK. We further demonstrate the compatibility of this technique with etched lithium niobate on insulator waveguides. The technique lifts performance limitations imposed by scattering as light transfers between photonic devices and optical fibers, paving the way for scalable integration of photonic technologies at both room and cryogenic temperatures.

Low-loss optical interfaces are crucial for optical information transport and processing across photonic and optoelectronic technologies. For instance, efficient light coupling between optical fibers and on-chip waveguides in photonic integrated circuits (PICs) has enabled numerous applications in optical/quantum interconnects [1, 2], optical switches [3], nonlinear optics [4], and quantum photonics [5, 6]. In particular, quantum information technologies rely on coherent control over interactions of single photons and qubits, which depend sharply on efficient transfer of quantum states of photons into and out of quantum PICs [7]. Significant progresses have been made in the chip-scale manipulation of photonic qubits for integrated quantum technologies [8], including all-optical quantum computing employing cluster states [9], quantum sensing [10], and quantum communication [11, 12]. However, scalable implementation of solid-state quantum computation [13] and communication [11, 12] platforms has been prevented by the long-standing challenge of efficiently and reliably transferring photons into and out of

---

[a)] Equal contribution

[b)] Author to whom correspondence should be addressed. Email:  machiel@amazon.com

quantum PICs at cryogenic temperatures, which are required to maintain high-fidelity quantum operations. This is extremely challenging because optical coupling interfaces made of heterogeneous materials experience thermal expansion mismatch as systems cool to the cryogenic temperatures, which imposes a fundamental tradeoff between coupling efficiency and stability at the interface of optical fibers and quantum PICs, limiting the effectiveness and reliability of these devices.

A variety of technologies are being studied to enable efficient, cryogenically-compatible photon transfer between optical fibers and PICs for scalable quantum technologies, but two technologies have emerged as dominant: Edge coupling and grating coupling. Edge coupling between optical fiber and on-chip waveguide is accomplished by guiding an on-chip waveguide to the edge of the sample and launching the optical mode into free space, from where it can be collected into a carefully aligned optical fiber. This technique has demonstrated high coupling efficiency with less than 0.5 dB of loss [14] but is limited by its sensitivity to spatial alignment accuracy, which can be disrupted by the thermal expansion mismatch during thermal cycles and mechanical vibrations [15, 16]. Thus, in-situ optical alignment is typically required to achieve relatively high coupling efficiency at cryogenic temperatures [16-19]. Recently a cryogenically-stable optical packaging for diamond nanophotonic devices without in-situ alignment has been demonstrated [20]. However, in this demonstration coupling loss between the optical fiber and the quantum device is limited to roughly 10 dB at visible wavelengths. For more established material platforms such as SiN and Si, the coupling loss is ~2.5 dB in the state-of-the-art cryogenically packaged devices [15]. On the other hand, grating couplers coherently scatter light from an on-chip waveguide into free-space using a periodically patterned grating such that it can be collected by an optical fiber. This technique has comparably higher tolerance to spatial alignment variations, and has been utilized to package nanophotonic devices with the required stability at cryogenic temperatures [21-23]. Because of the large contact area between fiber and device and the relatively relaxed alignment tolerance, the optical coupling of packaged fiber-grating interface remains stable over many thermal cycles from room temperature (300 K) to 7 mK in a dilution refrigerator. Unfortunately, grating couplers generally suffer from narrow optical bandwidth and limited coupling efficiency due to optical mode mismatch and grating directionality, resulting in a roughly 5 dB coupling loss per facet from an optical fiber to a silicon optomechanical crystal at telecom wavelengths [22].

Our work builds on the less established technology of adiabatic coupling [24], which utilizes a smoothly transitioning interface between tapered optical fibers and tapered waveguides to adiabatically match optical modes between these two devices. Using active alignment inside cryostats, this technique has demonstrated a low coupling loss of ~ 0.2 dB between optical fiber and nanophotonic waveguide with broad bandwidths [12, 24-27]. This exceptionally low loss makes it suitable for quantum applications where photon loss between quantum devices can quickly deteriorate performances [2, 7, 12]. Previously, in-situ alignment with complex nano-positioning stages and cryo-compatible imaging systems was required in this platform to compensate for thermal expansion mismatch, vibration induced shifts, and power-dependent instabilities at the optical coupling interface during thermal cycles, limiting the number of optical channels and increasing the system complexity and footprint. In order to eliminate the in-situ alignment process, here we demonstrate highly efficient and permanent packaging of the optical interface between a tapered optical fiber and diamond nanophotonic devices with a coupling loss of <1 dB per facet. The packaged nanophotonic device is cryogenically stable, surviving more than 5 thermal cycles down to 77 K with an efficiency change of less than 0.15 dB. And it is resistant to sudden thermal shocks, high-power illumination, and mechanical vibrations, enabling gas deposition based tuning and laser back-tuning of the resonance wavelength of the nanophotonic cavity inside in a dilution refrigerator [12]. Gas tuning can thus be used for individual targeting of cavity resonances to quantum emitters, facilitating the operation of a nanocavity integrated quantum memory. This technique is reproducible and compatible with multiple PIC platforms, including suspended diamond waveguides and thin film lithium niobate devices.

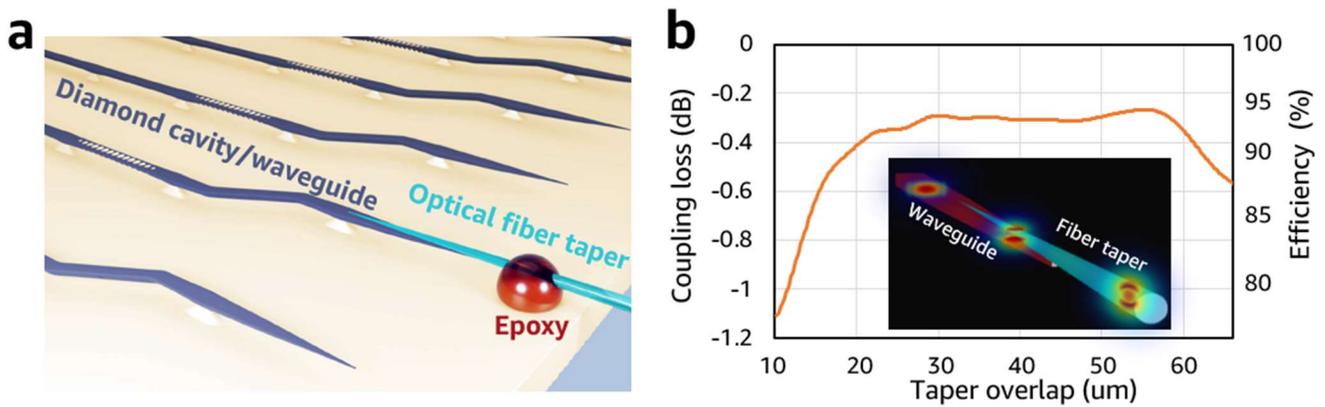

FIG. 1. (a) Illustration of the fiber-to-chip packaging concept. (b) Simulated coupling loss as a function of the taper overlap of an adiabatic interface between a tapered waveguide and fiber taper. Inset: Optical mode fields at different locations along the coupling interface.

Fig. 1(a) shows the schematic of our packaging approach. For most of the demonstrations in this paper we choose as our target device a free-standing diamond waveguide optimized for efficient photon exchange with silicon vacancy color centers [12, 26, 27]. The diamond device has a photonic crystal cavity which acts as a retroreflector away from its cavity resonance, and a tapered waveguide end used to create an adiabatic coupling interface with a tapered fiber [25-27]. The waveguide taper narrows from the waveguide width, 500 nm, to 100 nm over 20 µm distance. The fiber taper is made from a single mode S630-HP optical fiber (3.5 µm core and 125 µm cladding diameters) by chemical etching in hydrofluoric acid [25]. Withdrawing the fiber from the etchant bath at a constant speed tapers the diameter of the fiber down to ~150 nm at the fiber tip with a taper angle of about 1.5°. In simulations the coupling loss between optical fiber and waveguide remains below 0.4 dB for a wide variety of alignment parameters, as shown in Fig. 1(b). Changes in the overlap between the fiber and waveguide tapers of less than 40 µm (out of a total overlap distance of less than 60 µm) cause coupling fluctuations of only 0.2 dB, and transverse translation only modestly impacts coupling efficiency as long as the tapered ends remain in contact (See supplementary material). This waveguide-fiber interface is facilitated by strong van der Waals interactions, ensuring that once contact is made it can easily be maintained.

The alignment and assembly setup shown in Fig. 2(a) consists of motorized translation stages for fiber alignment, an optical microscope for imaging, an optical coupling loss measurement system, an epoxy dispenser, and a UV curing station. Once the fiber is aligned and coupled to the optical device, as observed using optical images in Fig. 2(c) and verified using the measured reflection off the photonic crystal in Fig. 2(d), an epoxy (Norland NOA 86TLH) droplet is placed ~200 μm away from the waveguide-fiber interface to keep the tapers aligned. The epoxy is immediately UV cured to prevent flow and more epoxy droplets are placed further down the fiber to secure the pigtail, as shown in Fig. 2(b). We further UV cure the package for several hours to fully polymerize the epoxy. Once cured the sample can be handled, transported, installed in vacuum chambers, and cooled down without specialized handling. After initial optimization of this procedure, we successfully packaged 5 successive devices with no process failures, reaching coupling efficiencies limited primarily by diamond device properties.

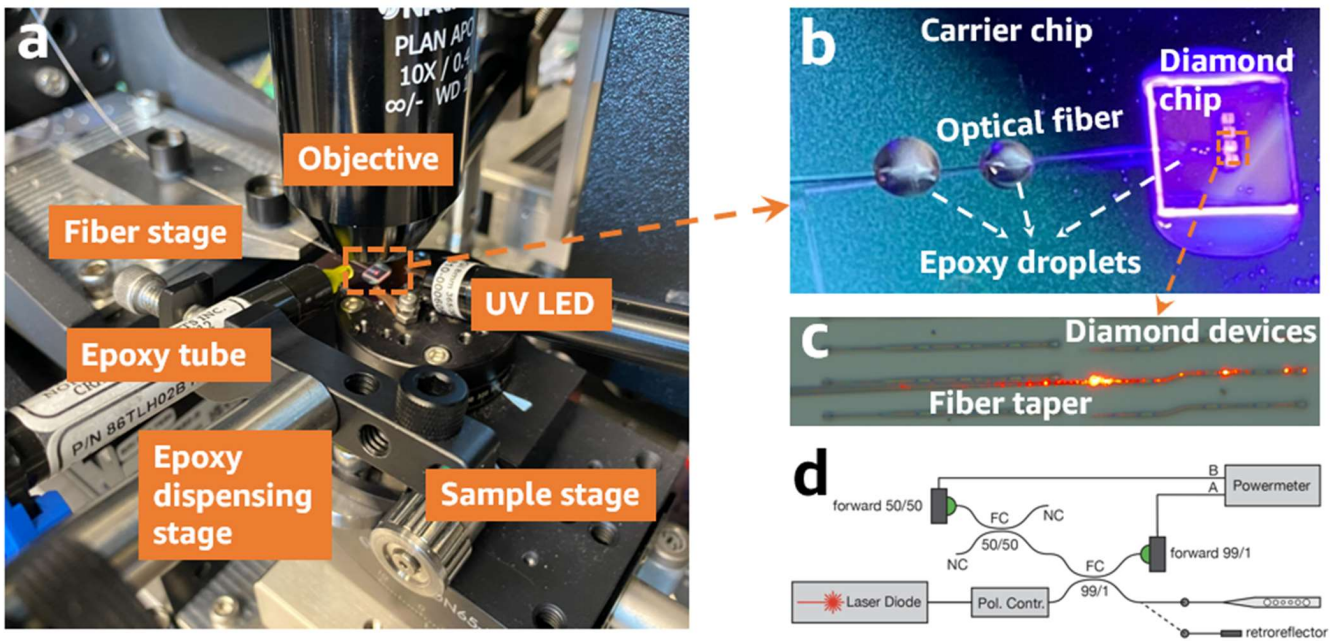

FIG. 2. (a) Customized alignment and assembly bench for fiber manipulation and gluing. (b) Fiber-packaged device under UV illumination showing a series of epoxy droplets. (c) Optical image of fiber-coupled diamond nanophotonic device, and (d) Fiber network for coupling loss measurements.

This procedure has a number of advantages over the state-of-the art photonic packaging techniques [20-23]. First, the measured change in insertion loss before and after gluing is negligible. This is because epoxy is far away from the point of contact and thus has a minimal impact on the coupling efficiency at the sensitive optical interface between fiber and waveguide. As such coupling efficiency is limited by the light scattering at the waveguide-fiber interface, not the packaging process added afterwards. Second, this technique enables multiple fibers to be attached to a single chip, as there is minimal crosstalk between the attachment and curing of multiple interfaces. Finally, this approach does not require the use of valuable chip-edge space, as the fiber arrives at an angle of approximately 8° and thus does not contact substrate at any point past the first epoxy droplet. This means that the chip edge real estate can be reserved for other optical and electronic interconnects and the density of interconnects on the chip can be increased.

To test the cryogenic stability of our packaged devices, we place each fiber-packaged device in a liquid nitrogen (LN2) flow cryostat as shown in Fig. 3(a). Because the coefficient of thermal expansion of materials drops at low temperatures contraction mainly occurs between room temperature and 77 K [22, 23, 28]. This means that cooling to LN2 temperatures provides a good indicator of the device performance at sub-Kelvin temperatures. We monitor the coupling efficiency in real time by connecting a cryostat feedthrough fiber to the fiber network in Fig. 2(d). To start a thermal cycle, we mount the sample inside the sample chamber on a copper pedestal using a conductive silver paste to ensure thermalization. The sample chamber is pumped with a turbo pump station to a base pressure of less than $10^{-4}$ mbar. Then we cool the sample at a rate of 3-4 K/min. To avoid optical tweezing of material freed from the cryochamber walls during cooldown, the laser is kept off until the temperature reaches 160 K. At this temperature the laser is turned on and 20 µW of power is sent to the waveguide and the intensity of the reflected light is monitored. We then keep the temperature constant at 77 K for an additional 30 minutes to allow the system to fully thermalize.

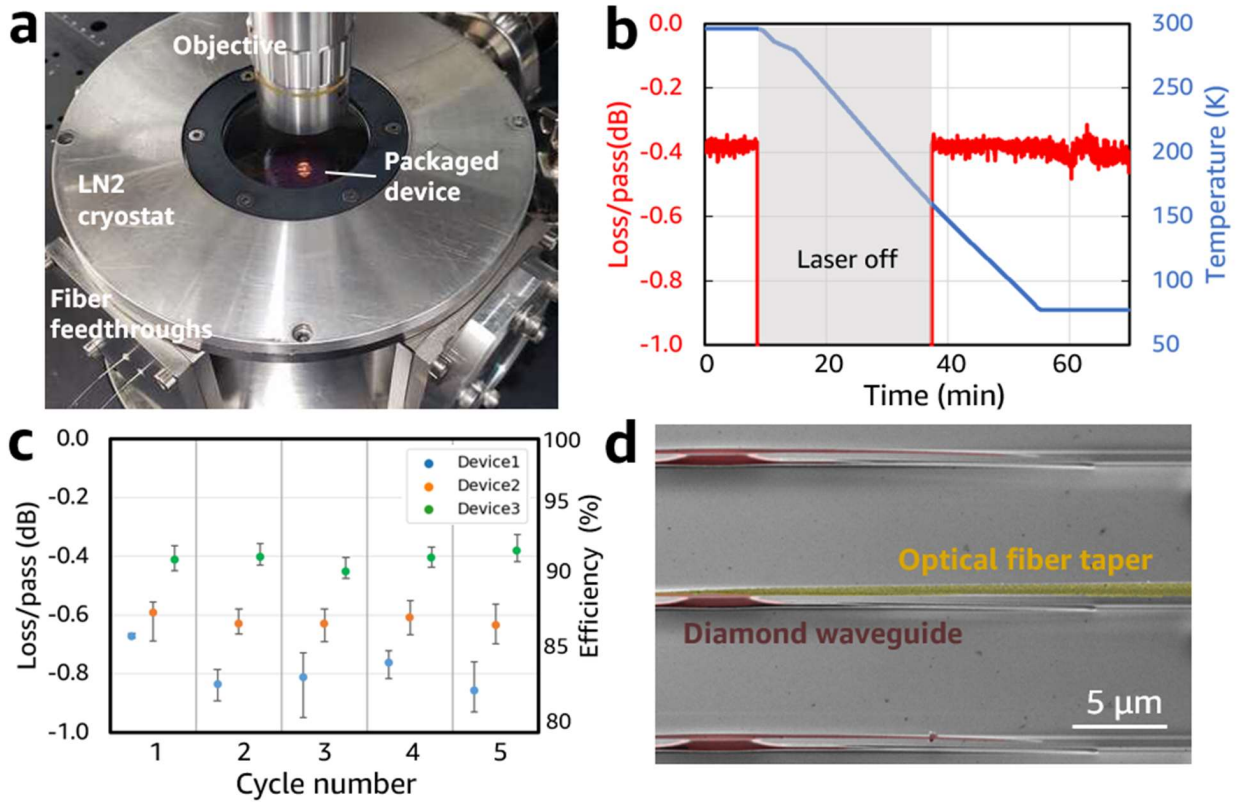

FIG. 3. (a) LN2 flow cryostat used for thermal cycling packaged samples down to 77K. (b) Measured loss and temperature as a function of time. Laser is turned off during initial stage of cooldown to avoid optical tweezing of material outgassed from cryostat chamber. (c) Statistical losses of 3 packaged devices over 5 thermal cycles. (d) SEM image of a packaged waveguide-fiber interface after 3 thermal cycles showing contact between tapered fiber and tapered diamond waveguide.

Fig. 3(b) shows the real-time coupling loss of a packaged device during one such thermal cycle. For this sample, once the laser is turned on the reflected power remained steady at ~83% of the reference reflected power (supplementary material). This quantity includes two waveguide-fiber interfaces, reflection off the photonic crystal, and propagation through roughly 150 μm of diamond waveguide. By taking the square root of this percentage we can thus establish a lower bound of our one-way fiber-waveguide coupling efficiency of 91%, or equivalently coupling loss of -0.4 dB. A small gradual decrease in coupling efficiency is observed as a result of temperature-induced stress impacting polarization of transmitted light. This can be corrected and the ~ 90% efficiency per facet can be recovered at low temperatures by changing the polarization of the input laser.

In Fig. 3(c), we document the statistical coupling loss at 77 K of three fiber-packaged devices as they pass through 5 thermal cycles. All samples show minimal performance degradation across 5 successive thermal cycles. The most heavily tested sample survived a total of more than 10 cycles without any observable changes in fiber alignment or coupling efficiency. Out of five tested samples, no failure was observed during thermal cycling. Fig. 3(d) shows a scanning electron microscope (SEM) image of a coupling interface after 3 successive thermal cycles. This confirms that the fiber taper remains firmly attached to the waveguide taper and the coupling interface is robust to the cryogenic environment.

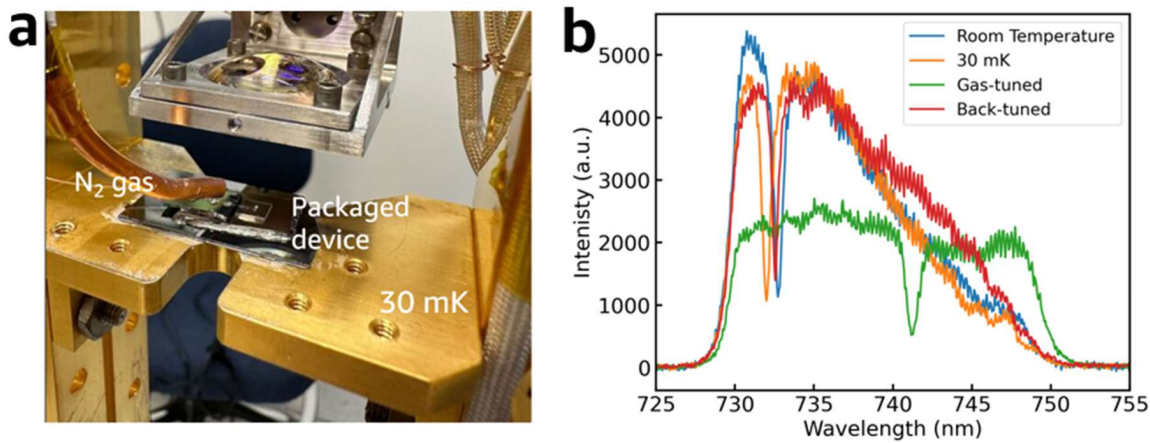

FIG. 4. (a) The packaged device with a fiber pigtail mounted in the mixing chamber of a dilution refrigerator. (b) Reflection spectra of the packaged diamond nanophotonic device at room temperature (blue), at 30 mK (orange), when gas-tuned (green), and back-tuned (red).

We further demonstrate our technique's cryogenic compatibility and utility for quantum optics experiments by cooling a sample inside a dilution refrigerator and tuning the optical resonances of photonic crystal cavities inside the waveguide by condensing nitrogen ($N_2$) gas onto the device. Gas tuning provides a mechanism to dynamically tune cavity resonance wavelength to overcome any inhomogeneity between fabricated devices and target emitters. This is done by injecting $N_2$ gas via a copper tube in Fig. 4 (a) near the fiber-packaged diamond device such that the nitrogen gas condenses on the sample surface. Deposited $N_2$ increases the refractive index of the device's environment and redshifts the cavity resonance. This technique redshifts all devices on the sample surface, at which point devices can be back-tuned individually by delivering high power laser via tapered fiber to locally evaporate $N_2$. This independent control over device resonances makes it possible to correct for fabrication imperfections and optimize each device resonance for spin-photon interactions. This experiment also demonstrates the waveguide-fiber interface's robustness to mechanical, chemical, and temperature perturbations which occur as hot gas lands on milli-Kelvin temperature devices and is subsequently boiled off using high powered lasers.

No degradation in coupling efficiency is observed as the device cools from room temperature to sub-Kelvin temperatures. Similarly, coupling efficiency remains steady as we carry out gas tuning by condensing nitrogen on the diamond surface tuning the cavity resonance by 10 nm. To tune the cavity back, we send 60 µW of laser to the device to induce local heating and evaporate deposited nitrogen, bringing the cavity back to within 1 nm of its original wavelength while preserving the coupling efficiency. Resonance spectra during the gas tuning experiment are summarized in Fig. 4 (b).

Finally, we test coupling to lithium niobate ridge waveguides and obtain better than -3.5 dB coupling per facet without modifying our procedure (See supplementary material for details). This efficiency was limited by the properties of the retroreflector on the lithium niobate chip being tested, and could be further improved by engineering lithium niobate geometry. Based on these results we believe this packaging approach could easily be adapted to operate with a wide variety of materials with low loss, and thus enable a variety of devices to be used in cryogenic, remote, or otherwise hostile environments.

Efficient collection of photons from single mode waveguides operating in cryogenic environments has been an outstanding problem in the quantum-optics community for many years. The cryogenically-compatible optical packaging technique demonstrated here provides an essential new tool for improved performance and large-scale deployment of classical and quantum optical devices to a wide variety of challenging environments. Packaged quantum optical devices improve their individual performance and makes possible the utilization of many chips inside the same cryogenic system or of multiple devices on a single chip (supplementary material). This is possible because the aligning and packaging processes used for this technique do not have significant chemical, optical, or mechanical crosstalk, enabling sequential or simultaneous packaging of many devices on a single substrate. Alignment can be accomplished using manually controlled micromanipulator stages — demonstrating the robustness of the technique to vibrations and small misalignment. These properties, combined with the small footprint, broad bandwidth, material agnostic nature, and high efficiency of our technique provide incentive for its further development.

**SUPPLEMENTARY MATERIAL**

See supplementary material for more details.

**ACKNOWLEDGMENTS**

We thank Nicholas Mondrik, Antia Lamas-Linares, Oskar Painter, Simone Severini, and Bill Vass for their operational support and technical guidance.

# SUPPLEMENTARY MATERIAL

## 1. Tapered fiber fabrication

To prepare tapered optical fibers, we followed a wet etching procedure outlined in [1]. In this process, bare single-mode optical fibers (Thorlabs S630-HP) were etched in a hydrofluoric acid (HF) bath capped with a thin layer of o-xylene to promote taper formation at the acid-oil interface. Before immersion in HF, fibers are cleaned in a 200 °C sulfuric acid solution. To control the taper angle, fibers were drawn out of the etching solution at a constant rate. This speed, and the initial fiber diameter, determine the final taper length and angle. We tuned this parameter to obtain a taper angle of less than 4° to enable adiabatic mode transition from optical fiber to nanophotonic waveguides. We automate this procedure by motorizing the fiber movements between the different solutions, enabling mass production of tapered fibers. The fiber tapers can be further engineered by varying the speed of withdrawal from HF solution or by varying the temperature of the solution while the process is underway, enabling fine grained control of the fiber taper shape.

## 2. Diamond photonic crystal cavity

Diamond waveguide devices consist of a triangular cross-section nanobeam waveguide with a photonic crystal cavity and a tapered end. Devices were designed and fabricated using methods reported in [2, 3]. The waveguide and photonic crystal cavity were designed to support transverse electric (TE) guided and cavity modes at ~740 nm, respectively. Simulated photonic crystal reflection spectra (Lumerical FDTD) show a reflection coefficient of better than 96% across the wavelengths measured, thus contributing less than 0.15 dB to the reported loss figure.

## 3. Coupling efficiency measurement setup

Fig. 2 (d) shows a fiber-based optical network used to measure the coupling efficiency of the waveguide-fiber interface. We send 730 nm laser light from a laser diode (Thorlabs LP730-SF15) through a polarization controller into the fiber network. Light is split in a 99:1 fiber beamsplitter with 1% of the light sent to the device and 99% measured as reference power (A) using a photodiode. The reflected light from the device is then split in a 50:50 fiber beamsplitter sending 50% of the light to another photodiode (B). The other 50% of light can be connected to a spectrometer for simultaneous measurement of cavity spectrum and coupling efficiency when a broadband supercontinuum source is used.

This setup enables accurate measurement of the reflection coefficient of the combined device and fiber interface while automatically compensating for fluctuations in the laser output power. The use of a single splice to connect the tapered fiber to the measurement setup introduces variation between packages of roughly 0.2 dB, thus driving the majority of the variation between packages observed in our measurement.

## 4. Coupling efficiency simulation

In order to investigate the alignment tolerance, we evaluate the fiber to waveguide coupling using Lumerical MODE solutions to simulate the coupling efficiency as a function of transverse shift and tilt of the fiber tip. The transverse offset only modestly impacts coupling efficiency 2% (~0.2 dB loss) when the misalignment is within ±100nm range of the ideal position, which can be easily achieved using commercially-available motorized translation stages. In addition, as long as tapered fiber tip is in contact with tapered waveguide and the titled angle is within ±2°, coupling efficiency remains above 88% (-0.55 dB loss). The simulations show that the coupling is insensitive to a certain amount of misalignment, which agrees well with experimental observations.

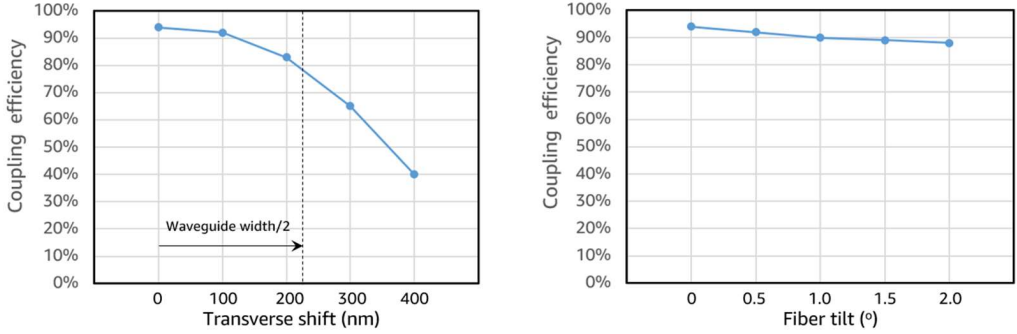

Fig. S1. Simulated coupling loss/efficiency as a function of the transverse offset, and tilted angle of the fiber tip.

## 5. Packaging thin film lithium niobate (TFLN) devices

To demonstrate the compatibility of the packaging technique with other platforms, we test the procedure for packaging TFLN ridge waveguides. A narrowband retroreflector and a suspended tapered waveguide were etched into TFLN on oxide to enable testing of adiabatic fiber coupling/packaging. In simulation per facet loss was less than 0.5 dB — in line with simulations of loss for diamond waveguides coupled in the same manner. In practice measured reflectance was roughly -6 dB for two facets - leading to an upper bound of per facet loss of -3.2 dB. This efficiency was limited by the properties of the retroreflector and the associated couplers on the TFLN chip being tested, which could be further improved by engineering device geometries. But even at this level of loss, the technique (with a lower level of loss, a smaller footprint, and a wider coupling bandwidth) is competitive with the state-of-the-art grating couplers for TFLN devices, in which in-situ alignment is required at cryogenic temperatures.

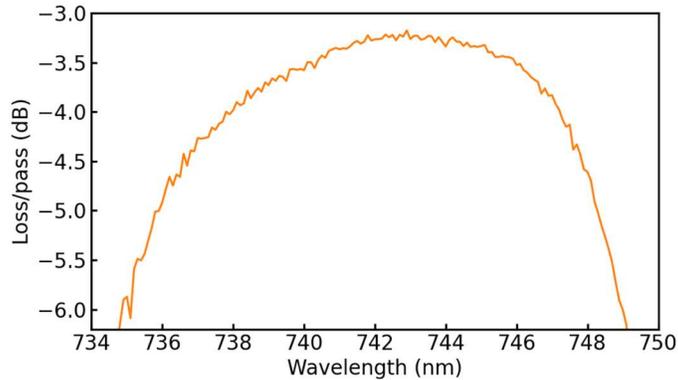

Fig. S2. Measured loss as a function of wavelength for TFLN devices.

Based on these results we believe this packaging technique could easily be adapted to operate with a wide variety of materials with low loss, and thus enable a variety of devices to be used in cryogenic, remote, or otherwise hostile environments.

## 6. Practical implementation for quantum memories

The packaging technique reported here will improve the performance and reduce the complexity of cryogenic quantum optics technologies, bringing them closer to deployment. In particular, packaging a tapered fiber to a device significantly reduces the complexity of cryogenic experiments used to operate diamond quantum network nodes [2] and state-of-the-art quantum transduction experiments [6]. As seen in Fig. S3 (a), using a packaged sample eliminates the need for a complex fiber coupling apparatus inside a dilution refrigerator. Specifically, this removes two sets of nanopositioning stages for fiber alignment, and the entire free-space imaging column from the system used to implement and image this alignment process. Removing these elements marks a significant step towards "plug-and-play" operation of quantum memories and enables the integration of multiple quantum memory devices inside a single cryogenic system, and thus their multiplexed utilization.

The multiplexing and high efficiency photon collection enabled by this technology will drive the performance of quantum repeater technology and photonic cluster state generation - key components of quantum communication and all optical quantum computation protocols. Fig. S3 (b) compares the performance of linear-optics measurement-device independent quantum key distribution (MDI-QKD) to upper-bounds for memory-enhanced MDI-QKD at different fiber-device coupling losses. Because the fidelity and rate of quantum communication drop at a superlinear rate as insertion loss increases [2] high efficiency photon collection plays a key part in the performance of quantum repeater protocols. Similar improvements are observed in cluster state formation where a single quantum emitter (coupled to a reservoir of nearby memories) [5] serves as a mechanism for generating interactions between photons which arrive asynchronously at the optical device in Fig. S3 (c). As insertion loss decreases the maximum achievable cluster state size grows — making possible increasingly sophisticated quantum computation protocols. Here again the maximum number of detected photons grows exponentially with reductions in insertion loss.

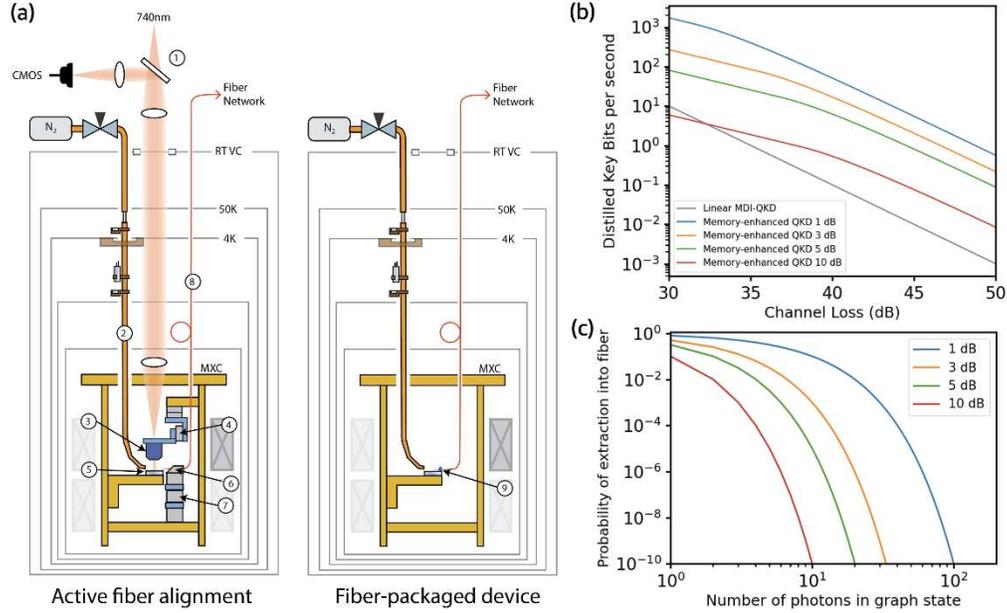

Fig. S3. (a) Footprint comparison of a gas-tuning experiment setup inside a dilution refrigerator with active fiber alignment apparatus versus fiber-packaging. 1: Frees-pace imaging column; 2: Gas tuning apparatus; 3: Objective; 4: Nanopositioners for objective; 5: Diamond chip containing waveguide devices; 6: Fiber mount; 16: Sample stage; 7: Nanopositioners for fiber; 8: Single-mode fiber S630-HP; 9: Fiber packaged device. (b) MDI-QKD communication rate using linear optics or memory enhanced quantum communication with device insertion loss limited by various efficiencies of optical fiber packaging. (c) Largest average cluster state that will be generated using quantum memories coupled to optical fiber with various levels of efficiency. Assumes a 10 MHz repetition rate for photon generation. In both (b) and (c) device performance scales superlinearly with reductions in insertion loss.

Memory-enhanced QKD of the kind demonstrated with SiVs [2] benefits heavily from a packaged fiber interface, which allows multiple quantum memories to be used on a single chip. Combined with wavelength division multiplexing, multiple quantum memories allow a linear scaling of qubit transmission rate. However, the performance of memory-enhanced MDI-QKD depends heavily on the efficiency of photon coupling from fiber to waveguide. The coupling efficiency affects the qubit rate both by adding additional effective loss to the incoming channels, but also, crucially, by determining what fraction of photons that have interacted with the memory arrive at the heralding detectors. Photons that interact with the memory but are lost before detection manifest as errors in the MDI-QKD scheme, which then require privacy amplification procedures to correct [4]. Privacy amplification reduces the fraction of arriving bits that can be used as key material.

To evaluate the performance of a QKD system using diamond nanophotonic cavities and glued fibers we calculate the effect of coupling efficiency $\eta$ on the total key rate of a memory-enhanced MDI-QKD system in the style of [2]. We assume that the base fidelity of the memory system (from memory qubit errors and finite spin-photon fidelity) is 95% with additional infidelity added by the probability of a photon interacting with the memory but being lost before the detectors due to $\eta$:

$$QBER = QBER_{base} + p(photon)(1 - \eta)$$

where N is the number of photon bins that are reflected from the memory during every attempt. We optimize N for every channel loss value to maximize the final key rate:

$$QKD\ Rate = r_S[Np(photon)\eta]^2$$

where $r_S$ is the fraction of raw bits remaining after privacy amplification. The results of this calculation are shown in Fig. S3(c).

Similarly, utilization of multi-photon quantum states for quantum information processing tasks such as one-way quantum communication and quantum computation require efficient optical coupling to quantum emitters [5]. For an N-photon state generated in a nanophotonic circuit, we can place an upper-bound on the probability of successfully coupling an N-photon state into a fiber of $\eta^N$, where $\eta = 1 - loss$ is the coupling efficiency from chip to fiber. Four example curves at 1, 3, 5, and 10 dB loss are plotted in Fig. S3(c), demonstrating the exponential decay in N-photon state detection probability as N increases. This illustrates the importance of chip-to-fiber coupling efficiency for experimentally generating, verifying, and utilizing photonic graph states for quantum information processing applications.